\documentclass[a4paper,12pt]{article}
\usepackage{graphicx,rotating,hyperref,slashed,amsmath,xcolor,amssymb,amsfonts,cite}
\makeatletter
\hypersetup{colorlinks,bookmarksopen,bookmarksnumbered,
linkcolor=blus,pdfstartview=FitH,urlcolor=rossos,citecolor=verde}
\allowdisplaybreaks

\def\lsim{\mathrel{\rlap{\lower3pt\hbox{\hskip0pt$\sim$}}
   \raise1pt\hbox{$<$}}}         
\def\gsim{\mathrel{\rlap{\lower4pt\hbox{\hskip1pt$\sim$}}
   \raise1pt\hbox{$>$}}}         

 \newcommand{\Mp}{M_{\rm Pl}}
 \newcommand{\bMp}{\bar{M}_{\rm Pl}}

\newcommand{\mio}[1]{}

\newcommand{\fig}[1]{~\ref{fig:#1}}

\definecolor{Gray}{gray}{0.95}

\newcommand{\sfrac}[2]{#1/#2}

\usepackage{multicol}
\usepackage{color}
\definecolor{rosso}{cmyk}{0,1,1,0.4}
\definecolor{rossos}{cmyk}{0,1,1,0.55}
\definecolor{rossoc}{cmyk}{0,1,1,0.2}
\definecolor{blu}{cmyk}{1,1,0,0.3}
\definecolor{blus}{cmyk}{1,1,0,0.6}
\definecolor{bluc}{cmyk}{1,1,0,0.1}
\definecolor{verde}{cmyk}{0.92,0,0.59,0.25}
\definecolor{verdec}{cmyk}{0.92,0,0.59,0.15}
\definecolor{verdes}{cmyk}{0.92,0,0.59,0.4}

\newcommand{\eq}[1]{~{\rm (\ref{eq:#1})}}

\newcommand{\GeV}{\,{\rm GeV}}

\def\circa#1{\,\raise.3ex\hbox{$#1$\kern-.75em\lower1ex\hbox{$\sim$}}\,}

\newcommand{\beq}{\begin{equation}}
\newcommand{\eeq}{\end{equation}}

\newcommand{\bea}{\begin{eqnarray}}
\newcommand{\eea}{\end{eqnarray}}
\newcommand{\be}{\begin{equation}}
\newcommand{\ee}{\end{equation}}
\font\tenrsfs=rsfs10 at 12pt
\font\sevenrsfs=rsfs7 at 10 pt
\font\fiversfs=rsfs5
\newfam\rsfsfam
\textfont\rsfsfam=\tenrsfs
\scriptfont\rsfsfam=\sevenrsfs
\scriptscriptfont\rsfsfam=\fiversfs
\def\mathscr#1{{\fam\rsfsfam\relax#1}}

\def\Lag{\mathscr{L}}

\def\circa#1{\,\raise.3ex\hbox{$#1$\kern-.75em\lower1ex\hbox{$\sim$}}\,}
\makeatletter

\def\hhref#1{\href{http://arxiv.org/abs/#1}{arXiv:#1}} 

\newcommand{\doi}[1]{\href{http://dx.doi.org/#1}{[link]}}

\def\hhref#1{\href{http://arxiv.org/abs/#1}{arXiv:#1}} 
 
\def\art{\@ifnextchar[{\eart}{\oart}}
\def\eart[#1]#2#3#4#5#6{{\rm #2}, {\em #3 \bf #4} {\rm (#6) #5} ({\em #1})}

\def\article{\@ifnextchar[{\earticle}{\oarticle}}
\def\oarticle#1#2#3#4#5#6{{\rm #1}, {\em ``#6''}, {\rm #2 #3 (#5) #4}}
\def\earticle[#1]#2#3#4#5#6#7{{\rm #2}, {\em ``#7''}, {\rm #3 #4 (#6) #5}  [\hhref{#1}]}
\def\hepart[#1]#2{{\rm #2, \em#1}}
\def\heparticle[#1]#2#3{#2, {\em ``#3''} [\hhref{#1}]}

\newcounter{alphaequation}[equation]
\def\thealphaequation{\theequation\hbox to
0.6em{\hfil\alph{alphaequation}\hfil}}
\def\eqnsystem#1{
\def\@eqnnum{{\rm (\thealphaequation)}}
\def\@@eqncr{\let\@tempa\relax \ifcase\@eqcnt \def\@tempa{& & &} \or
  \def\@tempa{& &}\or \def\@tempa{&}\fi\@tempa
  \if@eqnsw\@eqnnum\refstepcounter{alphaequation}\fi
\global\@eqnswtrue\global\@eqcnt=0\cr}
\refstepcounter{equation} \let\@currentlabel\theequation \def\@tempb{#1}
\ifx\@tempb\empty\else\label{#1}\fi
\refstepcounter{alphaequation}
\let\@currentlabel\thealphaequation
\global\@eqnswtrue\global\@eqcnt=0 \tabskip\@centering\let\\=\@eqncr
$$\halign to \displaywidth\bgroup \@eqnsel\hskip\@centering
$\displaystyle\tabskip\z@{##}$&\global\@eqcnt\@ne
\hskip2\arraycolsep\hfil${##}$\hfil& \global\@eqcnt\tw@\hskip2\arraycolsep
$\displaystyle\tabskip\z@{##}$\hfil
\tabskip\@centering&\llap{##}\tabskip\z@\cr}
\def\endeqnsystem{\@@eqncr\egroup$$\global\@ignoretrue} \makeatother

\definecolor{fiorentina}{rgb}{.5,0,.5}

\begin{document}
\centerline{CERN-PH-TH-2016-180}

\vspace{0.5truecm}

\begin{center}
\boldmath

{\textbf{\LARGE\color{magenta} On gravitational and thermal\\ corrections to vacuum decay}}
\unboldmath

\bigskip

\vspace{0.1truecm}

{\bf Alberto Salvio$^{a}$, Alessandro Strumia$^{a,b}$, Nikolaos Tetradis$^{a,c}$, Alfredo Urbano$^{a}$}
 \\[5mm]

{\it $^a$ CERN, Theoretical Physics Department, Geneva, Switzerland}\\[2mm]
{\it $^b$ Dipartimento di Fisica dell'Universit{\`a} di Pisa and INFN, Italy}
\\[2mm]
{\it $^c$ Department of Physics, University of Athens, Greece}\\[1mm]

\vspace{1cm}

\thispagestyle{empty}
{\large\bf\color{blus} Abstract}
\begin{quote}\large
We reconsider gravitational corrections to vacuum decay, confirming and simplifying earlier results
and extending them by allowing for a non-minimal coupling of the Higgs to gravity.
We find
that leading-order gravitational corrections suppress the vacuum decay rate.
Furthermore, we compute minor corrections to thermal vacuum decay in the SM by
adding one-loop contributions to the Higgs kinetic term, two-loop contributions to the Higgs potential
and allowing for time-dependent bounces.
\end{quote}
\thispagestyle{empty}
\end{center}

\setcounter{page}{1}
\setcounter{footnote}{0}

\tableofcontents

\newpage

\section{Introduction}
The measured Higgs mass lies 
close to the critical value above which the Standard Model (SM) Higgs potential is unstable at large field values.
In order to determine if the SM predicts that our universe is  stable or unstable,
several precision calculations were performed recently~\cite{1205.2893,1205.6497,1307.3536}, along with  
studies of gravitational corrections to vacuum decay~\cite{0712.0242,1508.05343,1512.01222,1601.06963,1601.07632,1603.07679,1606.00849,1606.07808} and of
cosmological implications~\cite{0710.2484,1301.2846,1308.2244,1403.6786,1404.5953,1407.3141,1404.3699,1404.4709,1409.5078,1506.07520,1505.04825,1607.00381}.
Also, the vacuum-decay formalism has been scrutinized~\cite{1604.06090}, and
better measurements of the top mass (the most unknown relevant parameter) are being planned,
from electroweak data, flavour data, LHC data,  and possibly new colliders~\cite{1508.05332}.
We contribute to this effort by addressing two concrete issues.

Concerning the vacuum decay rate, 
we show in section~\ref{Grav} that analytical techniques for including 
gravitational corrections at leading order in the inverse Planck mass~\cite{0712.0242}
provide correct results, contrary to the criticism of two recent papers~\cite{1601.06963,1606.00849}.
We extend and simplify the results of~\cite{0712.0242}.

Concerning the thermal tunnelling rate in the early universe, in section~\ref{T}
we extend previous calculations that included the one-loop thermal potential~\cite{termici},
by adding one-loop thermal kinetic terms (section~\ref{der}), two-loop thermal masses (section~\ref{2loop})
and allowing for time-dependent bounces (section~\ref{largeT}).

In section~\ref{concl} we present our conclusions. 
%
%

\section{Gravitational corrections to SM vacuum decay}\label{Grav}
Coleman and De Luccia developed a formalism 
for studying vacuum decay taking gravity into account~\cite{deluccia}.
However, the full theory of quantum gravity is unknown:
gravity is only known at the leading order in a low-energy expansion in inverse powers of  $M_{\rm Pl}$.
Thereby, the authors of~\cite{0712.0242} proposed a simple semi-analytical approximation that captures
the leading gravitational correction to vacuum decay.
The authors of~\cite{1601.06963,1606.00849} performed brute-force numerical computations of gravitational corrections
in Einstein gravity, and claim that the result of~\cite{0712.0242} is not valid.
We show that the original result in~\cite{0712.0242} is correct by 
providing further details on how it is obtained; 
we simplify the analytical expressions of~\cite{0712.0242}
and validate them through correct numerical computations.
We also generalize~\cite{0712.0242} to the case of a 
non-minimal coupling between the Higgs and gravity.

\subsection{The low-energy approximation}
We consider the Euclidean Einstein-Hilbert-Higgs action
\beq S =  \int d^4x \sqrt{g} \left[ \frac{(\partial_\mu h)(\partial^\mu h)}{2} + V(h)  -
\frac{\cal{R}}{2\kappa}- \frac{\cal{R}}{2} f(h)
\right],\label{eq:EH}
\eeq
where ${\cal R}$ is the Ricci scalar,
$\kappa = 1/\bar M_{\rm Pl}^2 = 8\pi G$ with $\bar M_{\rm Pl}=M_{\rm Pl}/\sqrt{8\pi}$, $M_{\rm Pl}\approx 1.22\times 10^{19}\,{\rm GeV}$.
For the moment we assume that the potential $V(h)$ and $f(h)$ are generic functions of the scalar field $h(x)$.
We allow for a generic non-minimal coupling to gravity $f(h)$, 
extending the formalism of~\cite{deluccia,0712.0242}. We
introduce an O(4)-symmetric Euclidean ansatz for the bounce $h(r)$ and for its
geometry
\beq  ds^2 = dr^2 + \rho(r)^2 d\Omega^2, \label{ansatz-metric} \eeq
where $d\Omega$ is the volume element of the unit 3-sphere.
On this background $S$ becomes
\beq S = 2\pi^2 \int dr \rho^3\left[\left( \frac{h^{\prime 2}}{2} + V\right)-
\frac{\cal{R}}{2\kappa}- \frac{\cal{R}}{2} f(h)\right], \label{Sexact}
\eeq
where now ${\cal R}=-6(\rho^2 \rho''+\rho \rho^{\prime 2}-\rho)/\rho^3$ 
and
a prime denotes $d/dr$.  The equations of motion are
\beq
h'' + 3\frac{\rho'}{\rho} h' = \frac{dV}{dh} -\frac12 \frac{df}{dh} {\cal R},\qquad
\rho^{\prime 2} = 1 + \frac{ \kappa \rho^2 }{3 (1+\kappa f(h))}\left(\frac{h^{\prime 2}}{2} - V-3\frac{\rho'}{\rho} \frac{df}{dh} h'\right), \label{EqExact}\eeq
where the latter equation can be obtained from the $rr$ component of the Einstein equations.
The bounce action in eq.~(\ref{Sexact}) can be simplified using a
scaling argument analogous to that of~\cite{coleman}:
the bounce action is  stationary under 
the rescaling $g_{\mu\nu}\to s^2 g_{\mu\nu}$. When this
rescaling is implemented in eq.~(\ref{eq:EH}), evaluated for the solution of
the equations of motion, the action should have 
an extremum at $s=1$. This observation relates the different contributions 
to the total integral that get multiplied by different powers of $s$. 
In particular, it implies that 
the bounce action can be simplified to
\beq S = -2\pi^2 \int dr \rho^3\, V, \label{Sexact2}
\eeq
evaluated on the solution of eq.s~(\ref{EqExact}) with the boundary conditions appropriate for a bounce.
This solution can only be obtained numerically.

\bigskip
Following~\cite{0712.0242} we  include analytically 
the effect of gravity, assuming $ R M_{\rm Pl}\gg 1$, where $R$ is the size of the bounce, by performing a leading-order expansion in  the gravitational
coupling~$\kappa$:
\beq h(r) = h_0(r) + \kappa h_1(r)+{\cal O}(\kappa^2)  ,\qquad \rho(r) = r + \kappa \rho_1(r)+{\cal O}(\kappa^2) .\eeq
The action $S_0$ at the 0th order in $\kappa$ is simply the scalar action in the absence of gravity computed for $h=h_0$.\footnote{The action contains the curvature term
enhanced by negative powers of the Planck mass.
Its expansion $-\sqrt{g}{\cal R}/(2\kappa)= 3 (r^2 \rho_1')' +{\cal O}(\kappa)$ apparently produces an extra 0th-order term.
However, this total-derivative term gives no contribution to $S_0$ for a $\rho_1'$ that is regular in $r=0$ and falls off sufficiently fast as $r\rightarrow \infty$.}
The action expanded at leading order in $\kappa$ is
\beq S= S_0+  \Delta S_{\rm gravity}, \eeq 
with
\beq \label{eq:Gexp} 
\Delta S_{\rm gravity}=  \frac{6\pi^2}{\bMp^2} \int dr~  \left[r^2 \rho_1 \left( \frac{h_0^{\prime 2}}{2} + V(h_0) \right)
+(r \rho_1^{\prime 2} + 2 \rho_1 \rho_1' +  2 \rho_1r \rho_1'')+
rf(h_0) (r \rho''_1 + 2\rho'_1)\right].\eeq
Notice that $h_1$ does not appear in eq.\eq{Gexp}.
The general reason behind this is that  the Higgs field sources gravity, but gravity does not source the Higgs.
A simplification of the above expression 
is possible through arguments similar to the one that led to eq.~(\ref{Sexact2}).
The total action can be viewed as a functional of $\rho(r)$ and $h(r)$, minimized
for the solution of eq.s~(\ref{EqExact}). Rescaling
$\rho_1(r) \to s \rho_1(r)$ corresponds to shifting the solution of the
equations of motion  
by $(s-1) \rho_1(r)$ (notice that this variation vanishes at the endpoints).
The action 
must have an extremum at $s=1$. Applying this
argument to (\ref{eq:Gexp}), by rescaling $\rho_1(r) \to s \rho_1(r)$
and requiring that the $s$-derivative of the resulting expression vanishes
at $s=1$, relates the integrals of terms linear and quadratic in $\rho_1$.
It leads to 
\beq \label{eq:Gexp2}
\Delta S_{\rm gravity}=  -\frac{6\pi^2}{\bMp^2} \int dr~  \left(
r \rho_1^{\prime 2} + 2 \rho_1 \rho_1' +  2 \rho_1r \rho_1'' \right)=
 \frac{6\pi^2}{\bMp^2} \int dr~r \rho_1^{\prime 2} \ge 0, \eeq 
where the last equation is obtained trough an integration by parts.
$\Delta S_{\rm gravity}$ is manifestly positive.
%
%
Once $h_0$ is known, $\rho'_1$
is given by eq.~(\ref{EqExact}) expanded at leading order in $\kappa$:
\beq
\rho'_1 = \frac{r^2}{6} \left[\frac{h_0^{\prime 2}}{2} - V(h_0) - \frac{3}{r} f'(h_0)  h_0'
\right], \label{eq:rho1Eq}\eeq
where $f'(h_0)$ is the derivative of $f$ with respect to $h$ evaluated at $h=h_0$.
Inserting this expression in eq.~(\ref{eq:Gexp2}) gives the leading-order gravitational correction to the action. 
Only an integration is needed.

\subsection{Gravitational corrections in a toy model}\label{toy}


 \begin{figure}[t]
\minipage{0.5\textwidth}
 \hspace{-0.4cm}
  \includegraphics[width=.85\linewidth]{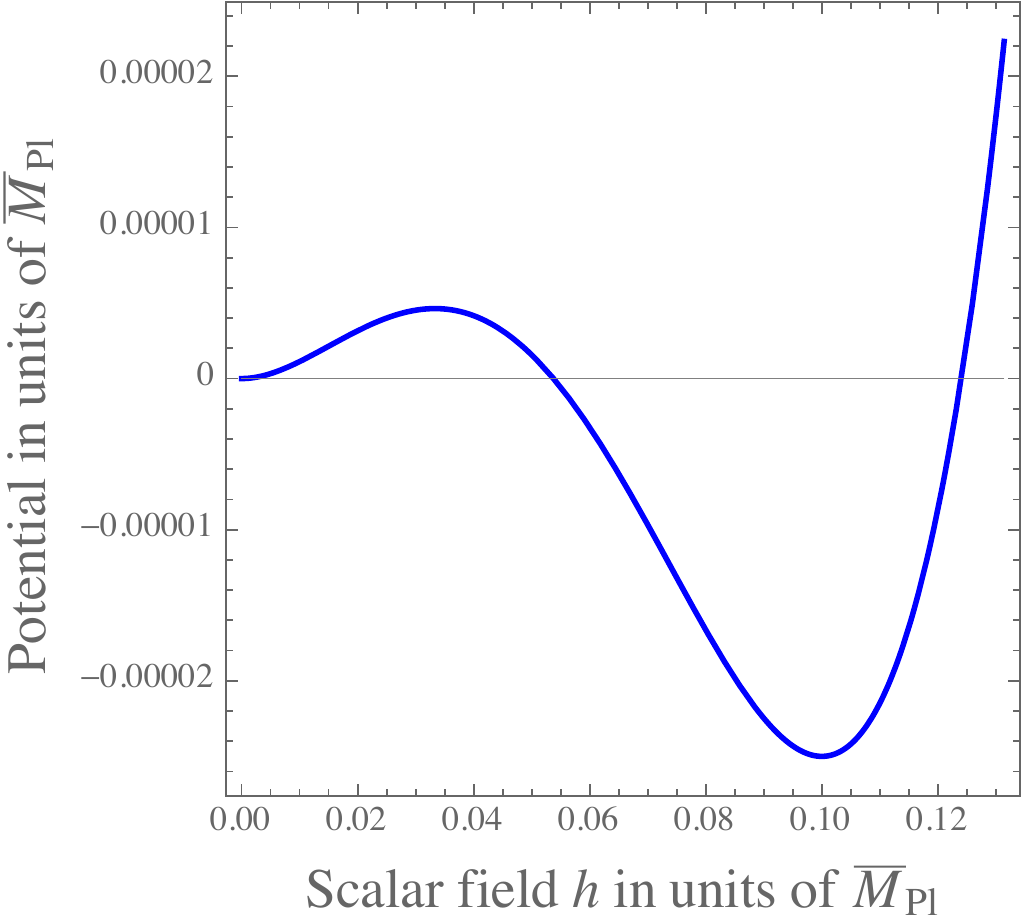}
\endminipage\hfill
\minipage{0.49\textwidth}
 \hspace{0.4cm}
  \includegraphics[width=.83\linewidth]{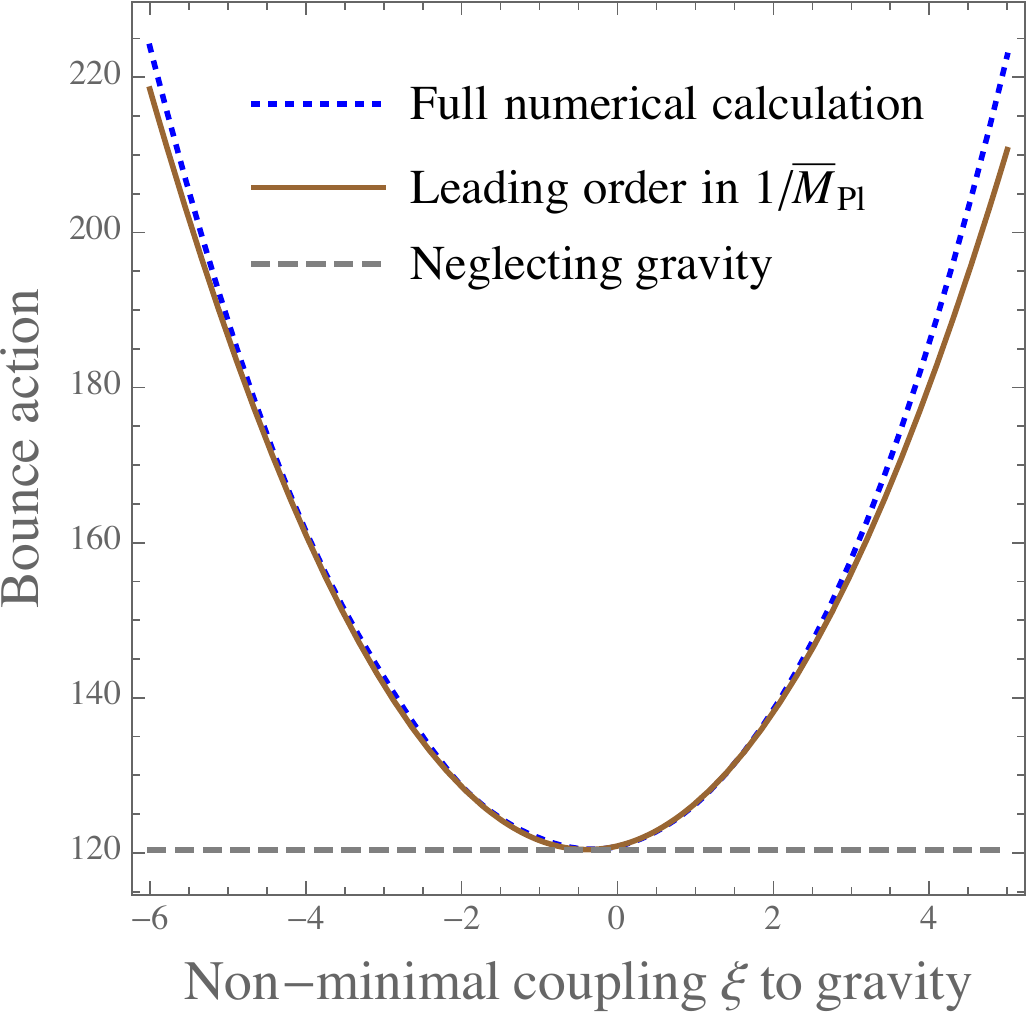}
\endminipage \vspace{.25 cm}
\caption{\em \label{BBB}
{\bf Left:} The potential (\ref{BBBpo}) for $g=3$, $b=1/3$ and $a= 0.05\, \bar M_{\rm Pl}$. {\bf Right:} The corresponding bounce action as a function of the non-minimal coupling $\xi$, comparing the full numerical result with the approximation at leading order in $1/M_{\rm Pl}$.
 } 
\end{figure}

Branchina et al.~\cite{1601.06963} performed a numerical analysis of vacuum
decay that resulted in the claim: 
``the output of~\cite{0712.0242} cannot be trusted and a fortiori cannot be used for comparison''.
We perform the comparison between numerical results and the semi-analytical approximation of~\cite{0712.0242} 
for gravitational corrections to vacuum decay, and find perfect agreement.
We consider the same quartic scalar potential studied in~\cite{1601.06963}\footnote{With respect to the conventions of \cite{1601.06963}, we have shifted the field so that the local minimum is located at $h=0$, and added a constant $V_0$ to the potential so that $V_{\rm B} (0) = 0$ at the false vacuum.} 
\beq V (h) = \frac{g^2}{4} \left\{\left[\left(h-a\right)^2-a^2\right]^2+\frac{4b}{3}\left[a\left(h-a\right)^3-3a^3\left(h-a\right)-2a^4\right]\right\} -V_0.\label{BBBpo}\eeq 
The left panel of fig.~\ref{BBB} shows the potential for a sub-Planckian choice of its parameters $g$, $a$ and $b$, and 
$f(h) = \xi h^2$.
In the right panel we show the bounce action as a function of $\xi$ in three cases: 
i) ignoring gravity;
ii) including gravity, with the perturbative approximation of eq.~(\ref{eq:Gexp2});
iii)  including gravity, performing a full numerical computation of eq.~(\ref{Sexact2}).

We see that the perturbative approximation reproduces the full numerical result.
For $\xi=0$ (the value considered in~\cite{1601.06963}) and the input values considered in fig.~\ref{BBB}, we find $S_0\approx 120.3$ and 
$S\approx 120.9$, which agrees with the perturbative approximation at the
per-mille level.
For larger values of $\xi$ gravity becomes stronger, and the perturbative expansion starts to break down, as expected.
We emphasize that a full numerical computation does not lead in an increase in precision, because
the semi-classical approximation too breaks down when gravity becomes strong.
Unknown quantum-gravity effects generically become relevant, as discussed in section~\ref{Pl}.

\subsection{Gravitational corrections to Higgs vacuum decay}\label{SM}
Rajantie and Stopyra~\cite{1606.00849} reconsidered the 
gravitational corrections to the 
vacuum decay rate in the Standard Model, concluding that:
``our numerical results are in conflict with~\cite{0712.0242}''.
We perform one more numerical computation, finding agreement with the analytical results of~\cite{0712.0242} and
clarifying the issues that led to the misunderstanding in~\cite{1606.00849}.



 \begin{figure}[t]
  \begin{center}
\fbox{SM with $M_h = 114$ GeV, $M_t = 173.34$ GeV, $\alpha_3(M_Z) = 0.1184$}\vspace{-0.4cm}
\end{center}
\minipage{0.5\textwidth}
  \includegraphics[width=.975\linewidth]{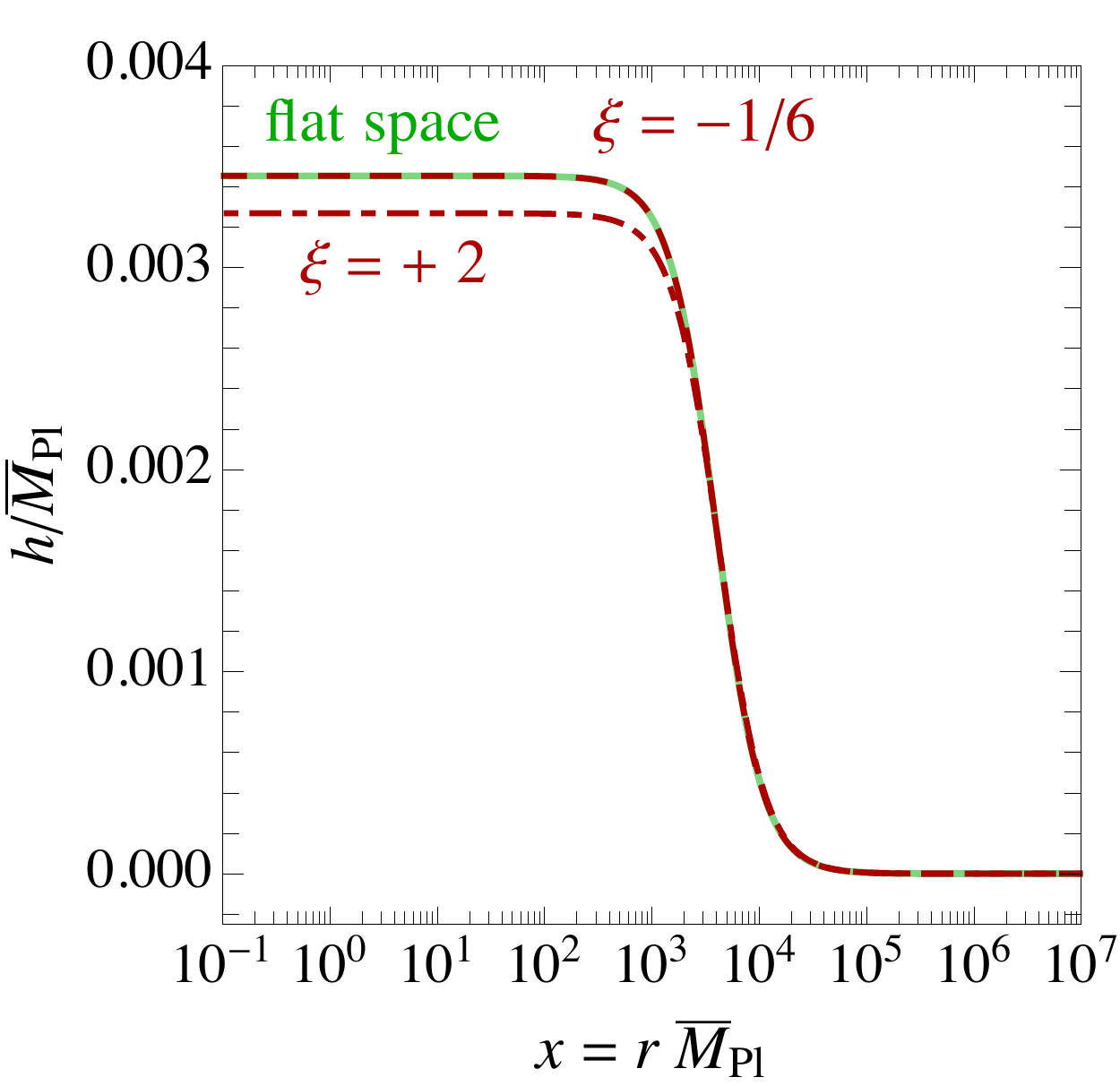}
\endminipage\hfill
\minipage{0.49\textwidth}
  \includegraphics[width=.975\linewidth]{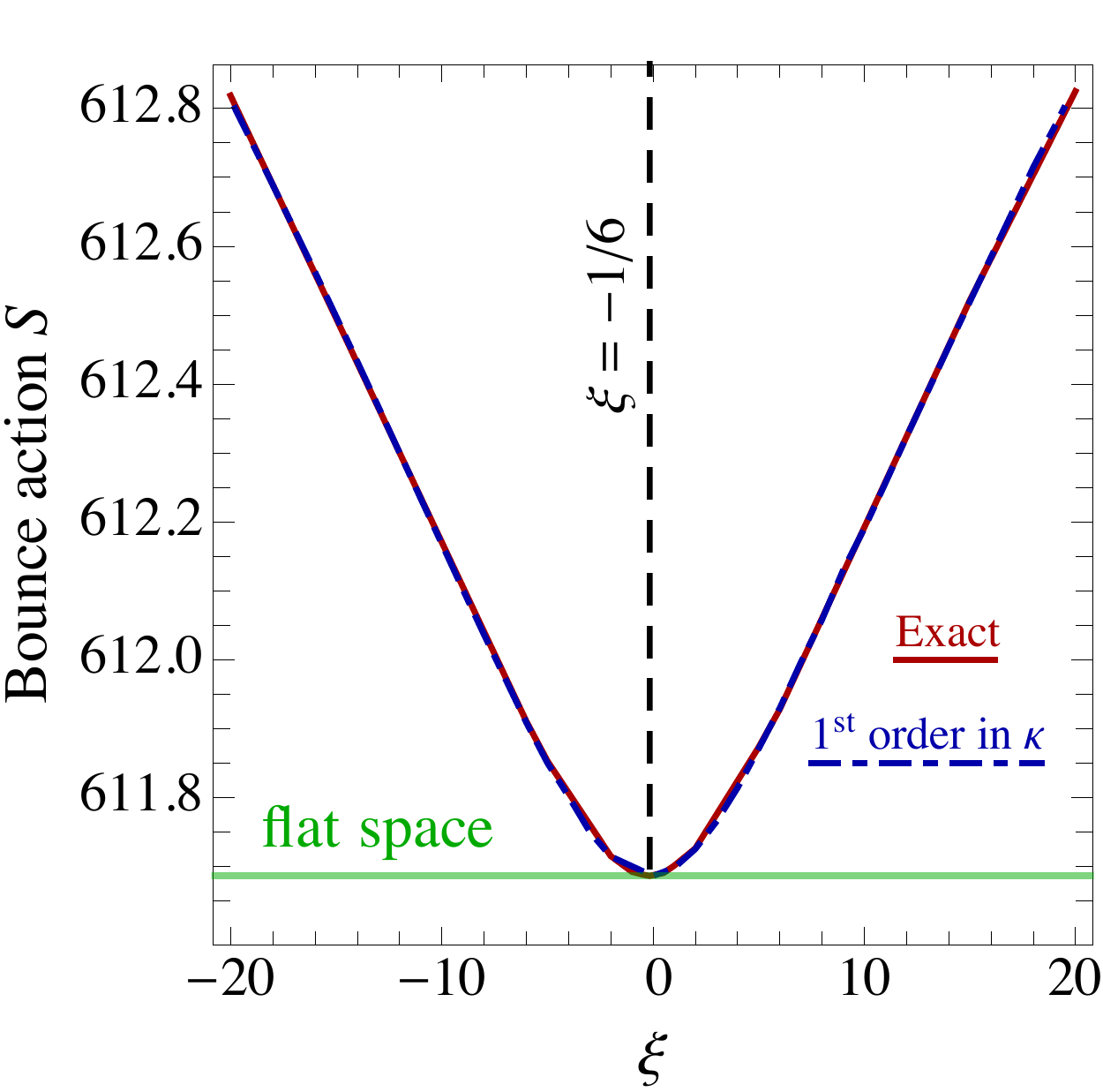}
\endminipage \vspace{.25 cm}
\caption{\label{fig:GravitySM114}\em SM bounce solutions for different values of $\xi$ (left panel), and their action (right panel).
We consider $M_h =114\GeV$, which is the value that saturates the meta-stability bound for the central value of the top mass.
The bounce is sub-Planckian, so that gravitational corrections can be 
computed perturbatively.}
\end{figure}

\smallskip

For the following we concentrate on a non-minimal coupling of the form
$f(h)=\xi h^2$. For a scale-invariant potential 
$V(h)=\lambda h^4/4$ with $\lambda<0$, and neglecting gravity, 
the bounce $h_0(r)$ can be computed analytically. It depends on an arbitrary scale $R$:
\beq \label{eq:h0}
h_0(r)= \sqrt{\frac{2}{|\lambda|}} \frac{2 R}{r^2+R^2}. \eeq
Quantum and gravitational corrections can be computed perturbatively by 
expanding around the solution of eq.~(\ref{eq:h0}).
Eq. (\ref{eq:rho1Eq}) becomes
\beq
\rho'_1 = 
\frac{8r^2 R^2}{3|\lambda|(r^2+R^2)^3}(1+ 6\xi). \label{rho1p}
\eeq 
Making use of eq.\eq{Gexp2} we obtain the final result:
\beq 
\label{eq:SSM}
 S =\min_R \left[ \frac{8\pi^2}{3|\lambda (\bar\mu)|}+\Delta S_{\rm quantum} +\Delta S_{\rm gravity} \right], 
\qquad \Delta S_{\rm gravity}=\frac{32\pi^2 (1+6\xi)^2}{45 (R \bMp \lambda)^2}.
\eeq
The gauge-invariant quantum correction $\Delta S_{\rm quantum}$ has been computed in~\cite{hep-ph/0104016} at one loop
in the $\overline{\rm MS}$ scheme. It
compensates for the RGE-scale dependence of $\lambda$,
such that one can conveniently choose the RGE scale $\bar\mu=1/R$.\footnote{The one-loop
calculation of the
decay rate basically amounts to substituting the tree-level action 
with the one-loop action.
The path-integral over all fluctuations has been computed 
in~\cite{hep-ph/0104016} up to
the last $\int d\ln R\, e^{-S(R)}$ integral over dilatations,
which is a higher-order effect because the SM tree-level action is scale-invariant.
The SM running of $\lambda$ fixes the intermediate value of $R$ that dominates the integral.
We adopt here the simple
Gaussian approximation, such that $\int d\ln R~e^{-S(R)}$ becomes $\min_R e^{-S(R)}$,
namely the least action principle.}
The gravitational correction at  leading order in $1/\Mp$, $\Delta S_{\rm gravity}$, agrees with~\cite{0712.0242}.
We included here the full quadratic dependence of $\Delta S_{\rm gravity}$  on $\xi$, going beyond the  linear order in $\xi$ computed in~\cite{0712.0242}.
Furthermore, $S$ indirectly acquires a different dependence on $\xi$ 
in view of the minimisation over $R$ dictated by eq.\eq{SSM}.

\smallskip

Fig.~\ref{fig:GravitySM114} demonstrates 
that the full numerical result agrees with the approximate expression.
We considered $M_h=114\GeV$, which is the value that saturates the  meta-stability bound
for the central value of the top quark mass. 
The bounce is sub-Planckian, such that gravitational corrections are small and can be reliably computed.%
\footnote{The authors of~\cite{1606.00849} justify
their criticism by claiming that no first-order correction $h_1(r)$ 
to the bounce 
with the correct boundary conditions $h_1'(0)=0$ and $h_1(\infty)=0$ exists.
While their calculation 
is technically correct, they miss the crucial physical point.
Indeed, they expand around the solution $h_0(r)$ of eq.\eq{h0}, which 
corresponds to the tree-level SM action that is scale-invariant and thereby
does not determine the scale $R$ of the bounce.
Adding only the effect of either gravity (operators with negative mass dimension)
or a Higgs mass term (operators with positive mass dimension)
results in either $R\to \infty$ or $R\to 0$:
namely the bounce no longer exists. 
The problematic $h_1$ is another manifestation of this issue.
In the real physical problem
the bounce exists because quantum corrections break scale invariance selecting an intermediate finite value for the bounce scale $R$,
 roughly given by the inverse scale that minimizes the running $\lambda$.
Therefore, 
the correct physical procedure is the one followed in~\cite{0712.0242}, and  summarised here in eq.\eq{SSM}:
compute the quantum corrections to the action as a function of $R$, 
and use them to determine  $R$. The gravitational corrections can then
be computed perturbatively. 
The solution for $h_1(r)$ is not needed in this calculation, but
can be computed from the quantum-corrected potential -- or any 
potential that fixes a scale for $h_0(r)$. The 
equation for $h_1(r)$ then has a solution that satisfies the correct
boundary conditions, thus 
resolving the issue raised in~\cite{1606.00849}.
}
Keeping instead $M_h$ at its experimental value and raising $M_t$ up to its meta-stability boundary $M_t\circa{<} 178\GeV$
again leads to a sub-Planckian bounce, with $h(0)\sim 0.1\bMp$.

\begin{figure}[t]
  \begin{center}
\fbox{SM with $M_h = 125.09$ GeV, $M_t = 173.34$ GeV, $\alpha_3(M_Z) = 0.1184$}\vspace{-0.4cm}
\end{center}
\minipage{0.5\textwidth}
  \includegraphics[width=.975\linewidth]{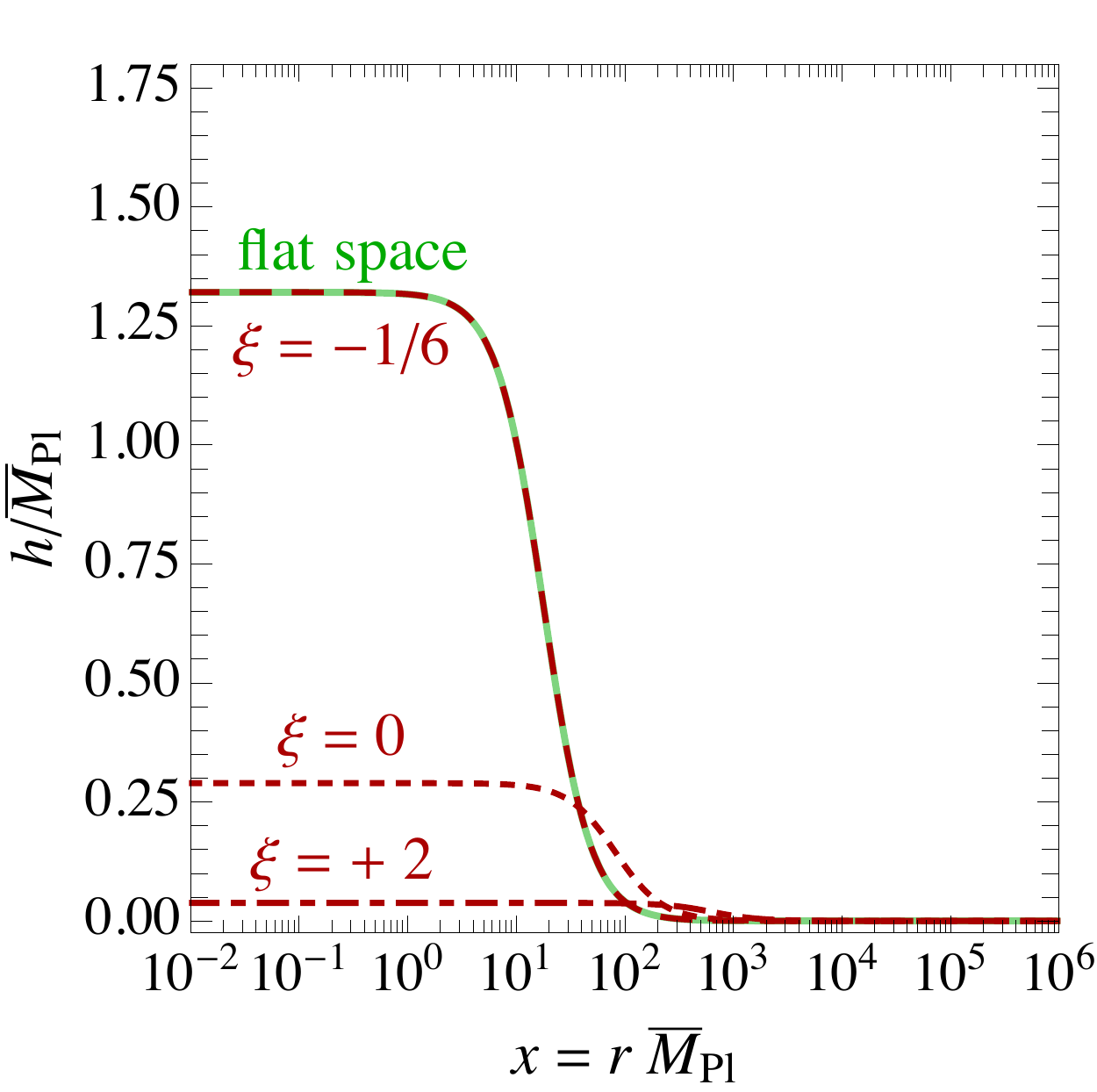}
\endminipage\hfill
\minipage{0.49\textwidth}
  \includegraphics[width=.975\linewidth]{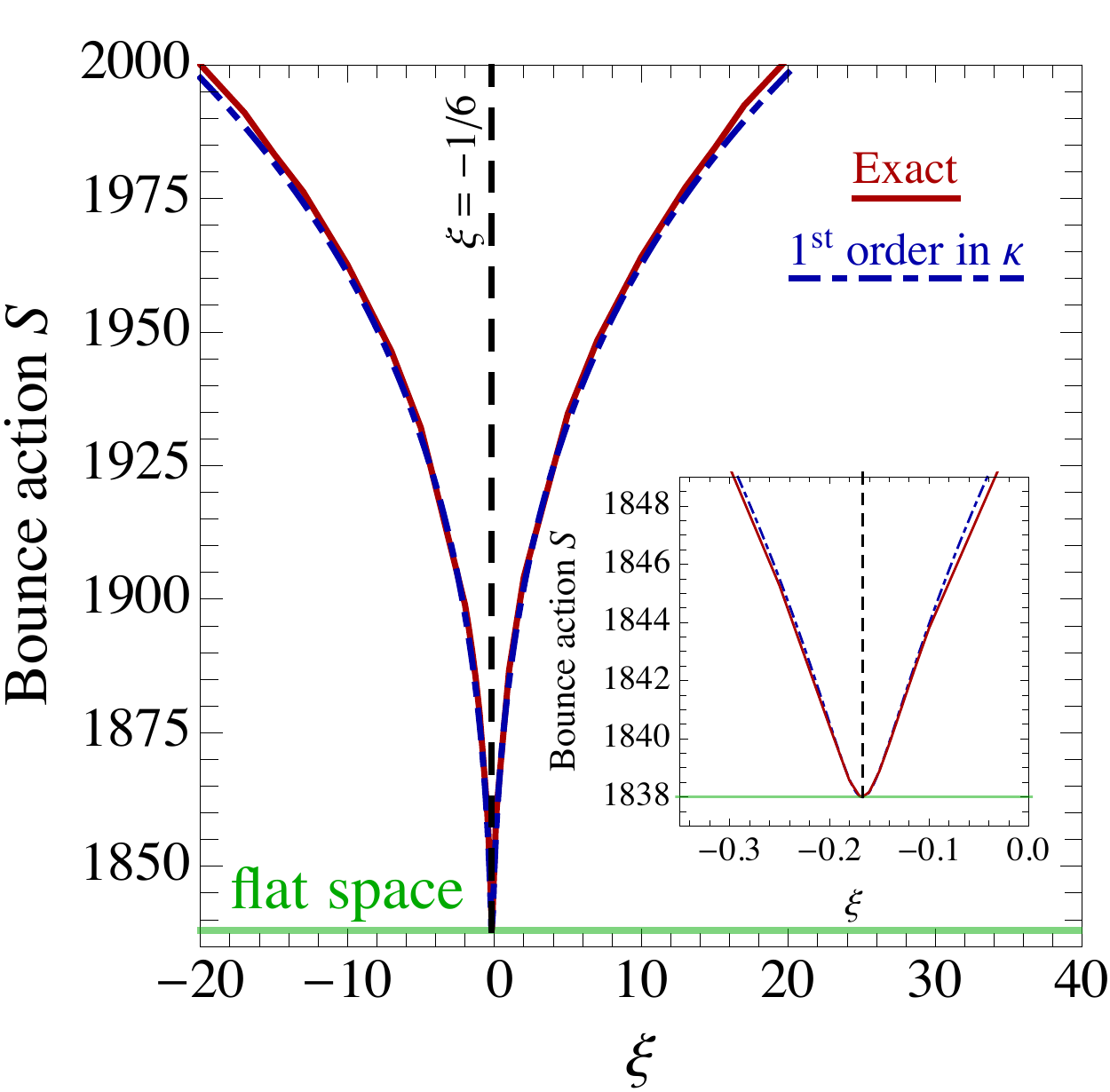}
\endminipage \vspace{.25 cm}
\caption{\label{fig:GravitySM}\em SM bounce solutions for different values of $\xi$ (left panel), and their action (right panel).
We consider here the best fit Higgs mass
$M_h =125.09\GeV$, for which the vacuum decay rate is negligibly small. 
For ease of visualisation we do not consider uncertainties due to higher order corrections.
}
\end{figure}

\begin{figure}[t]
 \begin{center}
  \includegraphics[width=.46\linewidth]{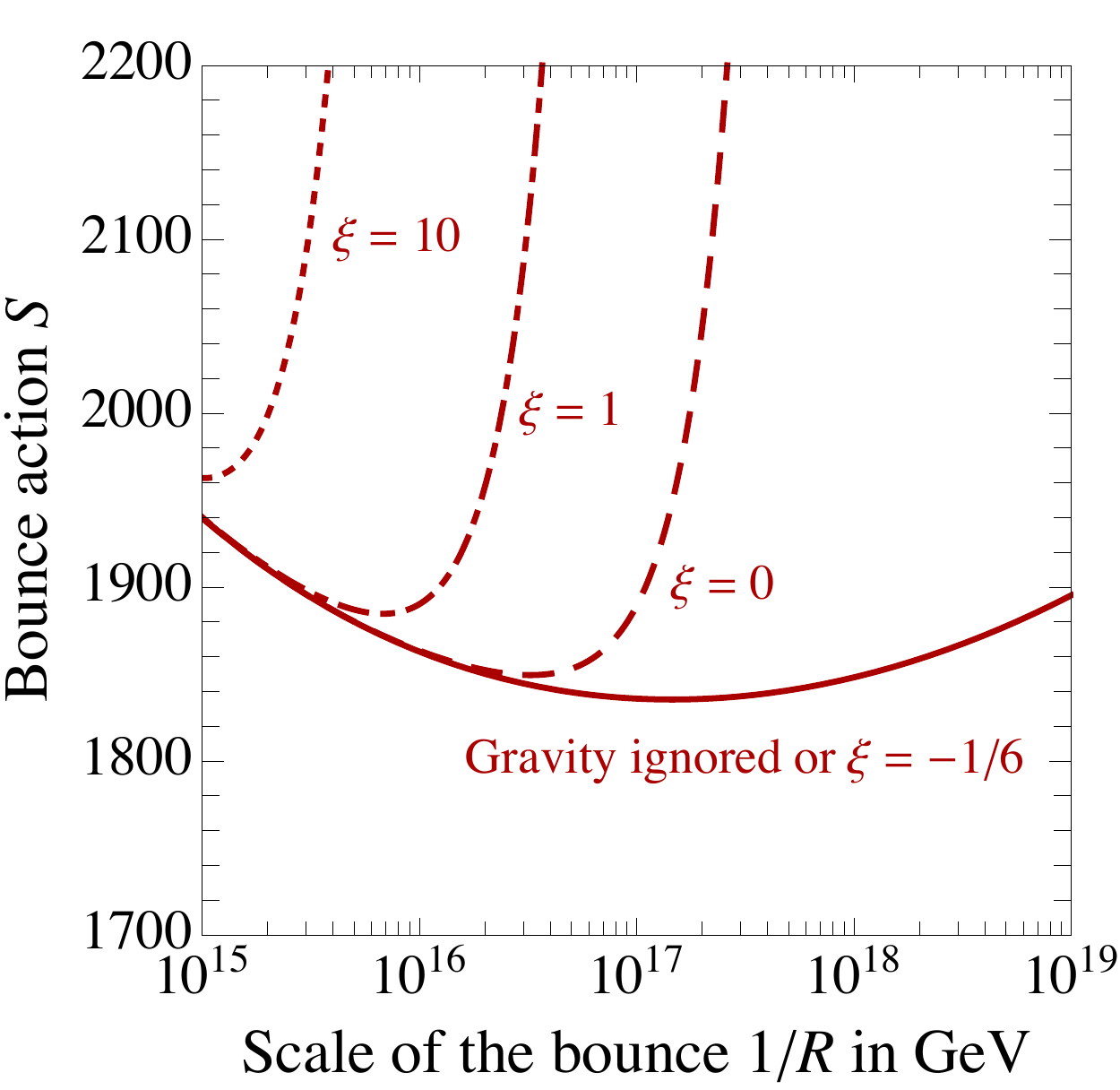}\qquad
    \includegraphics[width=.44\linewidth]{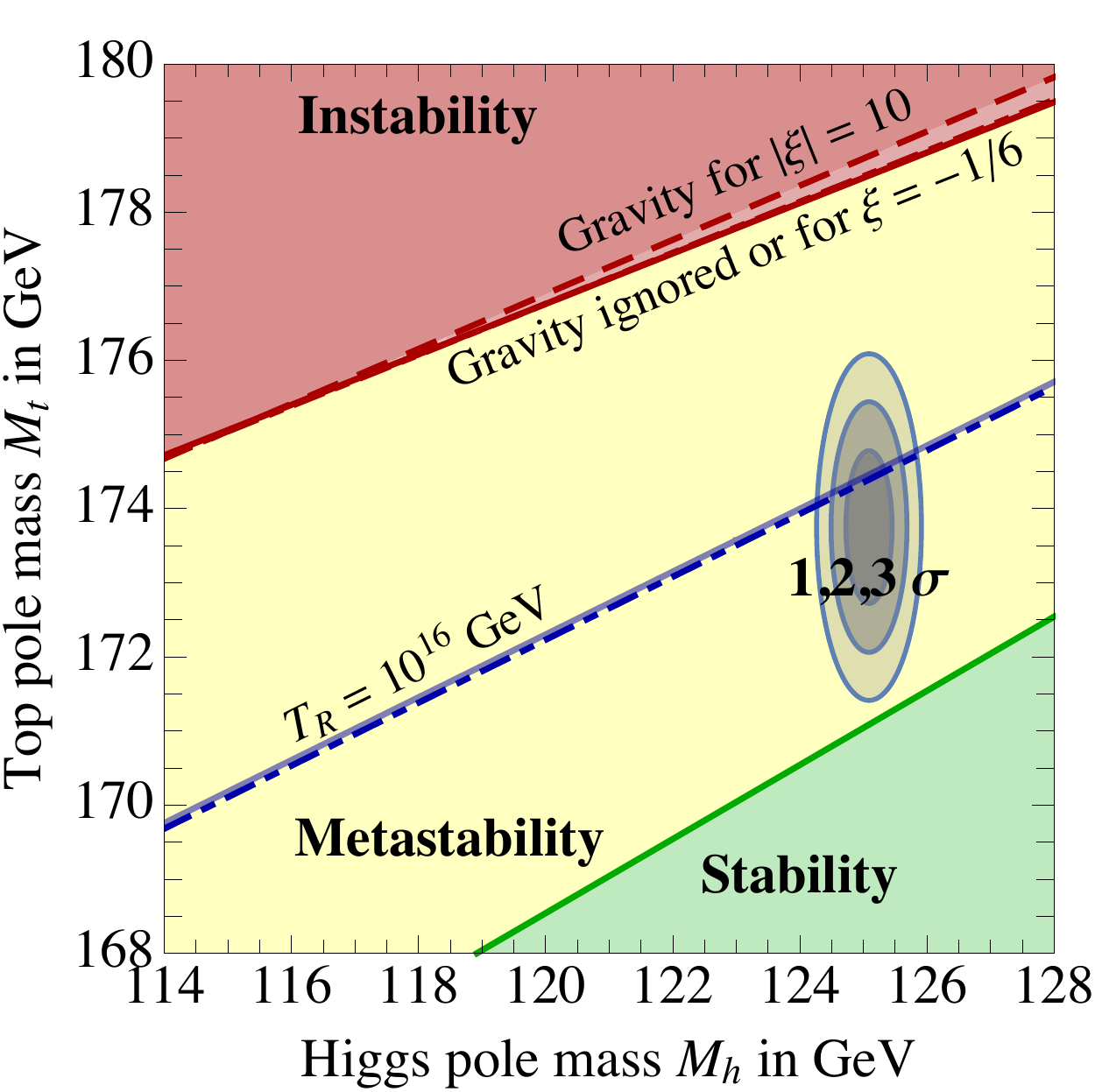}
\end{center}
\caption{\em {\bf Left}: bounce action as function of $R$  for $M_h=125.09\GeV$.
{\bf Right}: SM phase diagram for $\alpha_3(M_Z)= 0.1184$.
The continuous red line is obtained ignoring gravitational corrections or including them 
assuming the conformal value $\xi=-1/6$ of the Higgs coupling to gravity;
the almost coincident dot-dashed line assumes $\xi= 0$;
the dashed line assumes $|\xi|\sim 10$.
The ellipses show the measured values of the Higgs and top mass at $1,2,3\sigma$.
The middle blue lines are the bound from thermal tunneling, assuming
a reheating temperature of $10^{16}\GeV$.
\label{fig:PhaseDiagram}}
\end{figure}

In fig.~\ref{fig:GravitySM}  we consider the central value $M_h\approx 125.09 \GeV$,
which leads to a negligibly small vacuum decay rate dominated by a
bounce with  Planck-scale size, $h(0)\sim \Mp$.\footnote{For the sake of comparison, 
we explain the discrepancy between our fig.~\ref{fig:GravitySM} and the analogue plot in~\cite{1606.00849}:
they use the tree-level quartic potential $\frac14\lambda h^4$, 
while we use the 2-loop SM effective potential~\cite{1205.6497}. 
Both computations use 3-loop RGE running in the SM.
}
Naively, this is beyond the applicability domain of the low-energy expansion of~\cite{0712.0242}.
Nevertheless, the analytical approximation agrees well with the full numerical result because
approximate scale invariance combined with the
positivity of $\Delta S_{\rm gravity}$ implies that vacuum decay is dominated by bounces
with $h(0)\sim 1/R$ small enough not to be suppressed by gravity,
as illustrated in the left panel of fig.~\ref{fig:PhaseDiagram}.

The right panel of fig.~\ref{fig:PhaseDiagram} shows the SM phase diagram in the $(M_h, M_t)$ plane for $\alpha_3(M_Z)=0.118$.
We used our numerical code; 
the difference with respect to the analogous plot obtained from the analytical expression is
as small as unknown quantum-gravity effects.
We see that gravitational corrections have a minor effect:
the upper dashed line is obtained for $|\xi|=10$, and 
it differs by $\approx 0.5\GeV$ in $M_t$ from the dot-dashed line, obtained for $\xi=0$.
In turn, it is almost coincident with the continuous line, obtained either setting $\xi=-1/6$ or ignoring gravity.

This last feature is understood noticing that $\Delta S_{\rm gravity}$ vanishes for the conformal value $\xi =-1/6$.
This equality is not limited to the leading order in $1/\Mp$: the Fubini bounce of eq.~(\ref{eq:h0}),
together with the flat metric $\rho(r)=r$,
is an exact bounce solution of the full gravitational problem
for $\xi=-1/6$ and constant negative $\lambda$,
such that the bounce action is the same as in the non-gravitational case.
In particular, the last term in eq.~(\ref{EqExact}) identically vanishes.
Indeed, for $\xi = -1/6$, the Ricci scalar reduces to the simple form
$
\mathcal{R} = \kappa [
4V - h\,\sfrac{dV}{dh}
]$,
which vanishes for a scale-invariant potential $V = \lambda h^4/4$.
These properties can be also derived without any explicit computation from symmetry arguments:
for $\xi=-1/6$ the Higgs Lagrangian is conformally invariant; one can rescale the metric
so that any conformally flat metric, such as the one we consider in eq.~(\ref{ansatz-metric}), is equivalent to the flat metric. Thus, any solution of the $\lambda h^4$-theory on flat space-time is also a solution when gravitational effects are included.\footnote{The fact that the Einstein-Hilbert term breaks conformal invariance does not invalidate this conclusion. To show this consider conformal gravity (i.e.\ replace the Einstein-Hilbert term by the square of the Weyl tensor). The full theory is now conformally invariant. Any
solution of the $\lambda h^4$-theory on flat space-time is also a solution when gravitational effects are included. This implies in particular that the energy momentum tensor is zero for such a solution (recall that the Weyl tensor vanishes on flat space-time) and $\xi=-1/6$. So this configuration is also a solution of the Einstein-Hilbert-Higgs field equations.}

\subsection{Effects of new Planckian physics}\label{Pl}
Even if $h_0(0)\ll \Mp$,
Planck suppressed operators such as $|H|^6/\Mp^2$ and $|H|^2  |D_\mu H|^{2} /\Mp^2$
give extra corrections to the bounce action
of the same order as gravitational corrections:
at leading order in $1/\Mp$ they can be incorporated in $\xi$ through
field  redefinitions of the Higgs and of the graviton~\cite{0712.0242}.
Both $\xi$, as well as such effective operators, are unavoidably generated
when quantum corrections are added to the Einstein-Hilbert-Higgs action.
However, at higher orders an increasingly larger number of effective operators enters the game,
and the effective-theory expansion breaks down.

In order to compute if gravity suppresses or enhances vacuum decay,
one needs the theory of gravity, which is unknown. 
Assuming relativistic invariance,  general arguments suggest that such a theory
must either contain  an infinite number of positive-norm fields
(possibly resulting from some string theory) or a four-derivative graviton which includes one negative-norm component
(see~\cite{Stelle,1403.4226,1512.01237} for attempts to find a sensible quantum interpretation).

\medskip

The string solution suggests a complicated unknown
landscape of extra negative-energy AdS minima, and thereby
new contributions to vacuum decay.
As far as vacuum decay is concerned, the main implications of such a landscape are captured by adding
one new scalar $s$, possibly  with Planckian mass and decoupled from the Higgs.
Tunnelling along the $s$ direction opens a new channel for vacuum decay.
Its rate can be arbitrarily fast, independently of the mass of $s$.
This issue is orthogonal to SM vacuum decay:
Planck-scale physics cannot suppress sub-Planckian contributions to SM vacuum decay,
which can only be affected by new physics at lower energies.
In summary, calculations of the SM vacuum decay rate hold up to the  caveat 
`unless extra Planck-scale vacuum decay destroys the universe earlier',
analogously to how computations of the lifetime of SM particles hold up to the same obvious caveat,
which is conveniently left implicit.\footnote{The authors of~\cite{1307.5193} emphasize
that Planck-scale physics can give extra contributions to vacuum decay, but
proposing a specific example which relies on an uncontrolled expansion in $1/\Mp$:
an extra Planck-scale minimum in the Higgs potential obtained by adding terms $-h^6/\Mp^2$ and $+h^8/\Mp^4$.
}

\medskip

The second solution, which we refer to as ``agravity", gives more precise conclusions.
The  Euclidean Einstein-Hilbert-Higgs action is replaced by
\beq S =  \int d^4x \sqrt{g} \left[ \frac{(\partial_\mu h)(\partial^\mu h)}{2} + V(h)  -
\frac{\cal{R}}{2\kappa}- \frac{\cal{R}}{2}\xi h^2 -\frac{{\cal R}^2}{6f_0^2}+\frac{{\cal R}^2_{\mu\nu}-{\cal R}^2/3}{f_2^2}
\right],\label{eq:EHagr}
\eeq
where $f_0,f_2,\xi$ are dimensionless gravitational couplings, such that the theory is renormalizable. 
The term suppressed by $f_2$ gives rise to a ghost state which might admit a sensible physical interpretation~\cite{Stelle,1403.4226,1502.01334,1512.01237}.
In any case, this term does not contribute to the bounce action,
because it is the square of the conformally-invariant Weyl tensor, up to total derivatives,
and our background is  conformally flat.
The equation of motion for $h$ and the expression for ${\cal R}=-6(\rho^2 \rho''+\rho \rho^{\prime 2}-\rho)/\rho^3$ 
remain unchanged
while the equation for $\rho(r)$ becomes
\beq
\rho^{\prime 2} = 1 + \frac{ \kappa \rho^2 }{3 [1+\kappa (\xi h^2+2\mathcal{R}/3f_0^2)]}\left[\frac{h^{\prime 2}}{2} - V-\frac{\mathcal{R}^2}{6 f_0^2}  -3\frac{\rho'}{\rho} \left(2\xi h h' +\frac{2\mathcal{R}'}{3f_0^2}\right)\right] \label{EqExactagr}.\eeq
This equation can be obtained from the $rr$ component of the Einstein equations.
For our present purposes, it is convenient to ignore it, and rather close the system by adding the trace of the Einstein equations:
 \beq \bigg(\frac{\bMp^2+\xi h^2 }{2}-\frac{\Box}{f_0^2}\bigg)  \mathcal{R} = \frac{h'^2}{2}+2V +\frac{3\xi}{2}\Box h^2, \eeq 
 where $\Box$ is the covariant d'Alambertian.
This allows us to identify the main qualitative difference between agravity and Einstein gravity.
At energies much smaller than $f_0 \bMp$ the new $\Box$-term is irrelevant and one recovers the Einstein limit.
At larger energies, the $\Box$-term suppresses ${\cal R}$ with respect to the Einstein limit,
so that the gravitational correction to the bounce action saturates at
$|\Delta S_{\rm gravity}|\circa{<} \pi^2 f_0^2/\lambda^2$.
This means that gravitational corrections to SM vacuum decay can be ignored if $f_0$ is numerically small, as in~\cite{1403.4226}.
A negative value of $\xi$ (such that the Einstein term vanishes for $h= \bMp/\sqrt{-\xi}$) generates a
new vacuum instability.


\medskip

The only solid conclusion that one can draw from the above considerations is that new Planck-scale physics cannot cure the SM Higgs vacuum instability,  if such an instability appears much below $\bMp$.

\section{SM vacuum decay at finite temperature}\label{T}
The instability of the SM potential can also give rise to thermal tunneling in the early universe,
if it went through a hot enough phase (cosmological data only imply that the universe has been hotter than a few MeV).
The space-time probability density of thermal tunneling at temperature $T$ is 
given by 
\beq \gamma = \frac{d\wp}{d^4x} \approx  T^4 \bigg(\frac{S}{2\pi} \bigg)^{3/2} e^{-S}\eeq
where $S(T) = \int_0^{1/T} dt_E \int d^3x\, \Lag$ is the action of the thermal bounce at temperature $T$,
which is a solution to the classical equations of motion with periodicity $1/T$ in Euclidean time $t_E$.
The total cosmological probability of thermal tunneling up to today is obtained by 
integrating over the past light-cone
\beq 
\wp = \int dt \,dV  \gamma = V_0 \int  dt~ a^3~\gamma
\label{eq:tunrate}
\eeq
where $V_0=4 \pi (3.4/H_0)^3/3$ is the volume within the present horizon
and  $a$ is the Universe scale factor, equal to one today at $t=t_0$.
Using conservation of entropy to relate $a$ to $T$ we get
\beq \wp 
\approx \frac{ \sqrt{2}V_0}{\sqrt{\Omega_\gamma}H_0}  \frac{g_{*S0}}{g_*^{3/2}}  \int \frac{dT}{T} \bigg(\frac{T_0}{T}\bigg)^5 \gamma
\approx  117 \int \frac{dT}{T} \bigg(\frac{T_0}{T}\bigg)^5 \frac{\gamma}{H^4_0}
\eeq
where $T_0=2.7~{\rm K}$ is the present temperature,
$H_0\approx 67.4\,{\rm km/sec~Mpc}$ is the present Hubble rate,
$g_{*S0}=3.94$ the total number of effective degrees of freedom contributing to
the entropy after $e^+ e^-$ annihilation, and $g_*=106.75$ the number of SM degrees of freedom at $T$ 
much larger than the electroweak scale, when the thermal probability receives the dominant contribution. 
A small probability of thermal tunnelling $\wp\ll  1$ is roughly obtained if $S(T)\circa{>}206+\ln(\Mp/T)$ at any $T$
below the reheating temperature.


In the following we revisit computations of the thermal tunnelling rate adding 
three new effects to previous computations. 
In section~\ref{2loop} we include two-loop corrections to the thermal Higgs potential.
In section~\ref{der} we include one-loop derivative corrections to the thermal Higgs action.
In section~\ref{largeT} we explore time-dependent bounces.

\begin{figure}[t]
\minipage{0.5\textwidth}
  \includegraphics[width=.975\linewidth]{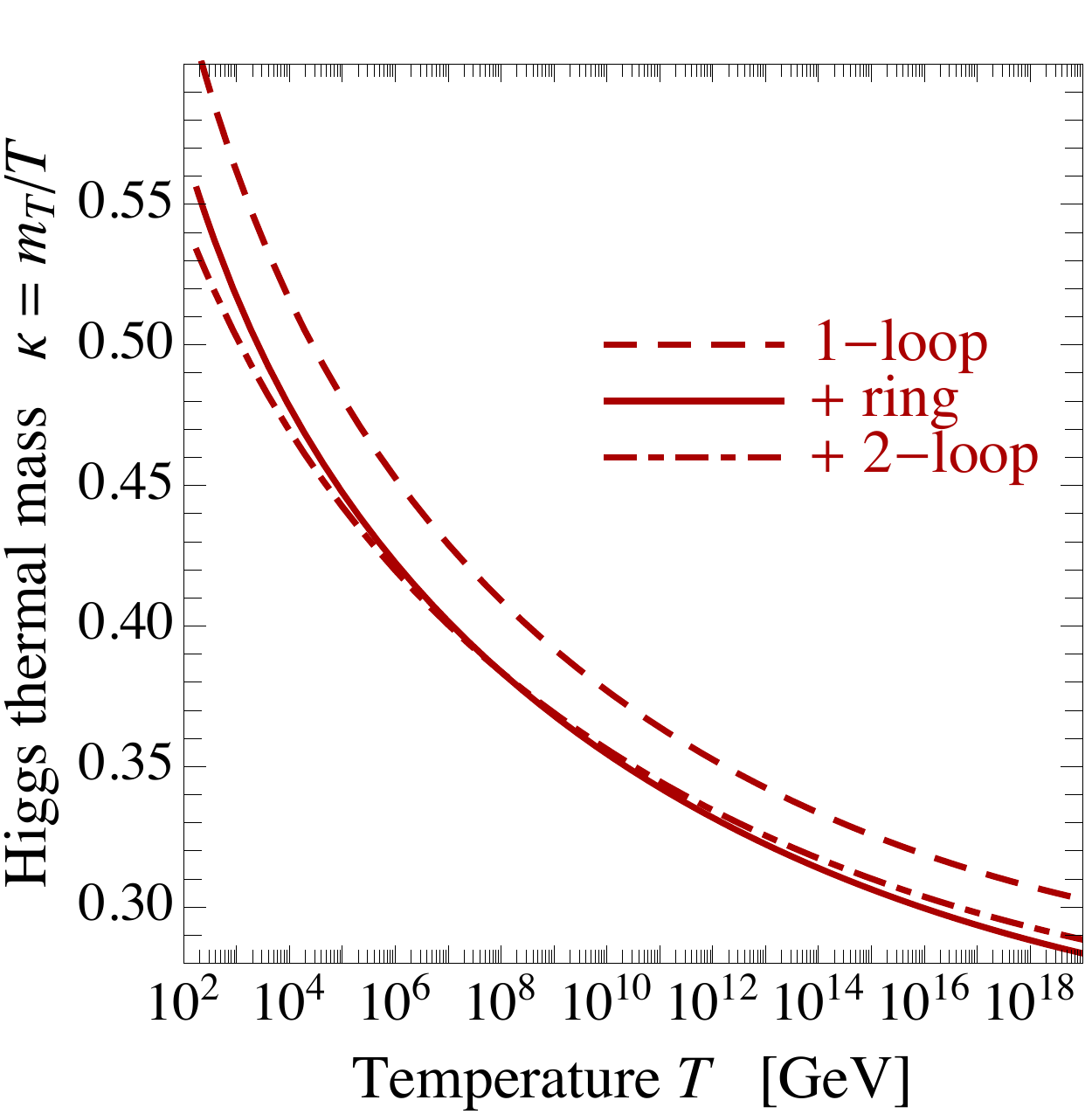}
\endminipage\hfill
\minipage{0.49\textwidth}
  \includegraphics[width=.975\linewidth]{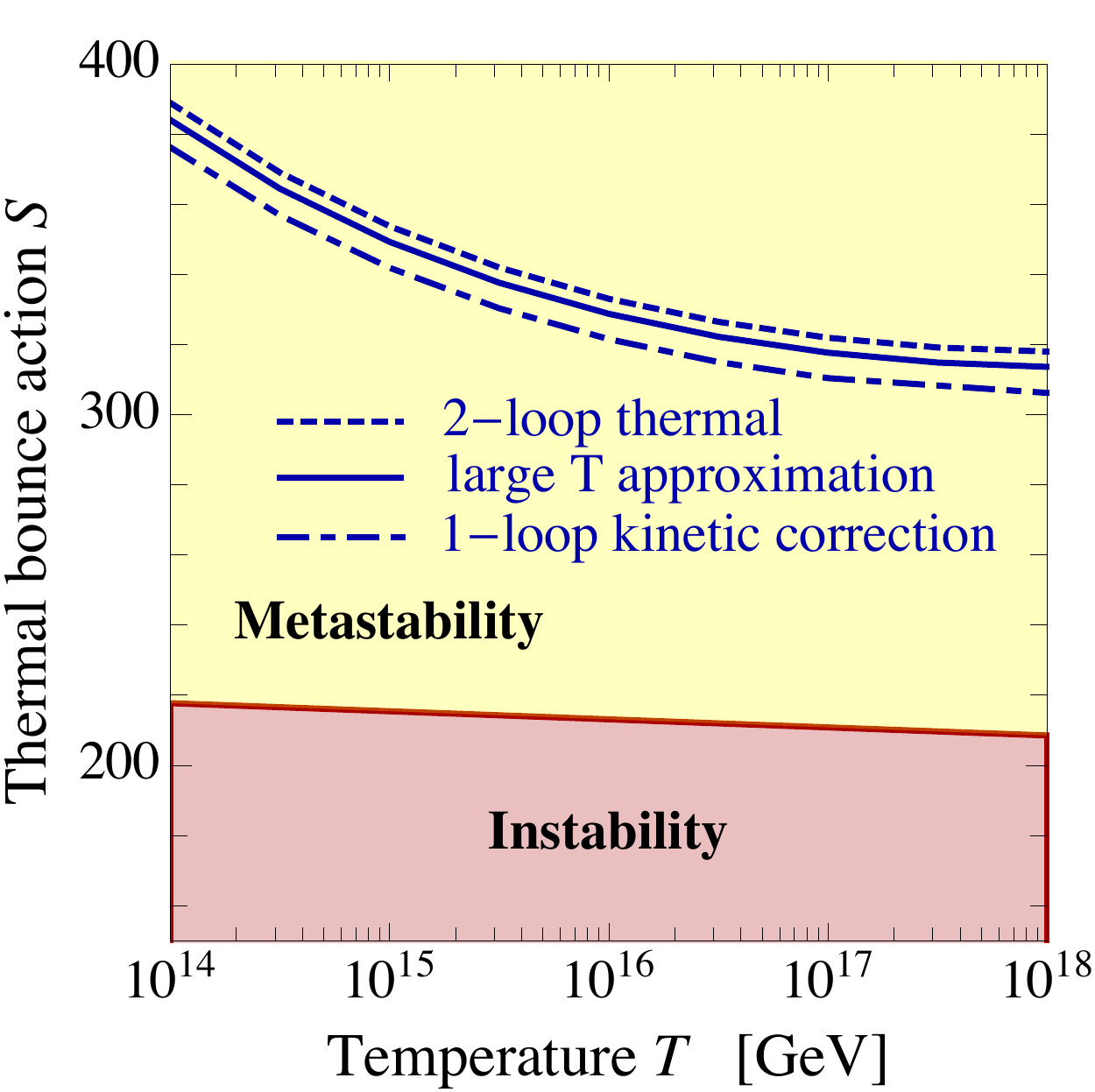}
\endminipage \vspace{.25 cm}
\caption{\em We consider the SM 
 for $M_h= 125.09\GeV$, $M_t= 173.34\GeV$,
 $\alpha_3(M_Z)=0.1184$.
{\bf Left}: Higgs thermal mass $m_T/T$ as function of the temperature, as precisely defined in eq.~(\ref{eq:PotThermal}),
computed adding higher-order corrections in the thermal loops.
{\bf Right}: action of the thermal bounce $S(T)$ computed with
the usual large-temperature approximation (solid curve), 
adding 2-loop thermal masses (dotted),
1-loop kinetic corrections (dot-dashed).
We also show the boundary between stability and meta-stability.\label{fig:EffectiveCoupling}
 }
\end{figure}

\subsection{Two-loop Higgs thermal mass}\label{2loop}
The temperature-dependent  effective potential can be expanded as
\begin{equation}\label{eq:PotThermalFull}
V_{\rm eff}(h,T) = V_0(h) + V_{\rm 1-loop}(h) + V_{\rm 2-loop}(h) + 
V_{\rm 1-loop}(h,T) + V_{\rm ring}(h,T)+ V_{\rm 2-loop}(h,T) \cdots~,
\end{equation}
where  the   first  three terms refer  to $T = 0$.  
To make the structure of the effective potential more transparent, a reasonable approximation is
\begin{equation}\label{eq:PotThermal}
V_{\rm eff}(h,T) \approx m_T^2(h) \frac{h^2}{2}   + \frac{\lambda_{\rm eff}(h)}{4} h^4.
\end{equation}
The  effective quartic coupling $\lambda_{\rm eff}$  is extracted from
the RG-improved effective potential at two-loop order and zero temperature.
The two-derivative Higgs kinetic term is canonically normalized, up to corrections not enhanced by large logarithms.
We write the Higgs thermal mass as $m_T^2 \equiv \kappa^2 T^2$ with
$\kappa^2 = \kappa^2_{\rm 1-loop}+\kappa^2_{\rm ring}+\kappa^2_{\rm 2-loop}$ and
\begin{eqnsystem}{sys:thermal}
\label{eq:ThermalMass}
\kappa^2_{\rm 1-loop} &=&
\frac{1}{16}g^{\prime 2} + \frac{3}{16}g^2 +\frac{1}{4}y_t^2 +\frac{1}{2}\lambda~,\\
\kappa^2_{\rm ring} &=&-
\frac{1}{16\pi}\sqrt{\frac{11}{6}}\left(g^{\prime 3} + 3g^3\right) -\frac{3\lambda}{8\pi}\left(
g^{\prime 2} + 3g^2 + 8\lambda +4y_t^2
\right)^{1/2}.
\end{eqnsystem}
Higher-order corrections to $\kappa$ are given in \cite{hep-ph/9212235,hep-ph/9403219},
and contain logarithmic factors that cancel the dependence on the RG-scale $\bar\mu$
of the lower-order terms, roughly
dictating that the running couplings in eq.~(\ref{sys:thermal}) are renormalised at $\bar\mu\sim T$.
We fix the residual RG-scale dependence setting $\bar\mu=T$;
$m_T^2$ acquires a logarithmic dependence on $h$, and
in the left panel of fig.~\ref{fig:EffectiveCoupling} we plot its value at the relevant scale $h=T$.
We see that the 2-loop contribution is small.



In the right panel of fig.~\ref{fig:EffectiveCoupling} we show
that including the 2-loop thermal mass gives a small correction to the bounce action, at the few $\%$ level.
This is consistent with the fact that the 2-loop correction to $\kappa$ is small
and that the bounce action is roughly proportional to $\kappa$ (if the full thermal potential is approximated through
a constant $m_T$ and a constant $\lambda_{\rm eff}$,
the bounce action is $S\approx 6.015 \pi \kappa/\lambda_{\rm eff}$~\cite{Arnold}).

\subsection{One-loop thermal correction to the Higgs kinetic energy}\label{der}
Various authors computed the one-loop thermal potential.
However, the bounce action receives comparable contributions from the kinetic part of the Lagrangian.
The computation of vacuum decay at $T=0$ has been performed including the full one-loop effective action~\cite{hep-ph/0104016},
which includes an infinite number of derivatives.
Performing similar computations at finite $T$ is more difficult:
we study here the impact of thermal corrections to the two-derivative Higgs kinetic term.

One-loop thermal corrections to derivative terms in the effective action at finite temperature were presented in~\cite{Moss:1985ve,Bodeker:1993kj}
and are of relative order $g^2/4\pi$.
We can focus on  spatial derivatives, because they receive the main correction in the large-$T$ limit and because
the thermal bounce is time-independent (see section~\ref{largeT}).
Such corrections can be written as
\be \Delta S= \frac12 \int_0^{1/T} dt_E \int d^3x\,Z_2(h,T)  \, ( \partial_i h)^2  \ee 
where $i$ runs over spatial coordinates.
For the SM at large temperature, $Z_2$ is given by
 \begin{eqnarray}
Z_2(h, T) & \approx & \frac{T}{4\pi}\left\{
\frac{\lambda^2 h^2}{4}\left[
\frac{3}{m_h^3(T)} + \frac{1}{m_{\chi}^3(T)}
\right] - \frac{4g^2}{3}\left[
\frac{1}{m_{\chi}(T) + m_W} 
\right]+
\right.\nonumber \\
&-&  \left.\frac{2g^2}{3c_{\rm W}^2}\left[
\frac{1}{m_{\chi}(T) + m_Z} 
\right]  +\frac{g^2 m_W^2}{12}
\left[
\frac{1}{2m_{W_L}^3(T)} + \frac{5}{m_W^3}
\right]  +
\right. \label{eq:dK}\\
&+& \left.
\frac{g^2 m_Z^2}{24}
\left[
\frac{c_{\theta}^2}{2{m}_{Z_L}^3(T)} + \frac{5}{m_Z^3}
\right]
+
\frac{g^2 m_Z^2}{24}\left[
\frac{s_{\theta}^2}{2{m}_{\gamma_L}^3(T)} + \frac{8s_{\theta}c_{\theta}}{({m}_{Z_L}(T)+ {m}_{\gamma_L}(T))^3}
\right]
\right\}.\nonumber
\end{eqnarray}
Thermal masses 
$m_{i}^2(T) = m_i^2 + \kappa_i^2 T^2$ for
$i = h,\chi,W_L,W_T, Z_T, \gamma_T$
can be computed in terms of   the usual field-dependent zero-temperature mass $m_i$, 
and of~\cite{Quiros:1994dr}
\begin{equation}
\kappa_h  = \kappa_\chi = 
\frac{3g^2 + g^{\prime 2}}{16} + \frac{\lambda}{2} + \frac{y_t^2}{4},\qquad
\kappa_{W_L} =  \frac{11}{6}g^2,\qquad
\kappa_{W_T}=\kappa_{Z_T}=\kappa_{\gamma_T} =0.
\end{equation}
The masses $m_{Z_L}$ and $m_{\gamma_L}$ are the eigenvalues of the thermal mass matrix~\cite{Quiros:1994dr}
\begin{equation}
\left(
\begin{array}{cc}
{m}_{Z_L}^2(T)  & 0   \\
0  &  {m}_{\gamma_L}^2 (T) 
\end{array}
\right) = R 
\left(
\begin{array}{cc}
m_Z^2 + \Pi_{Z_L Z_L}(T)  &   \Pi_{Z \gamma_L}(T)    \\
 \Pi_{Z \gamma_L}(T)   &  \Pi_{\gamma_L \gamma_L}(T)
\end{array}
\right)
R^T~,~~~R =
\left(
\begin{array}{cc}
c_{\theta}  & -s_{\theta}   \\
 s_{\theta} &    c_{\theta}  
\end{array}
\right)~,
\end{equation}
where $R$ is the matrix that rotates the mass eigenstates at $T=0$ into those at $T\neq 0$. This matrix is defined in terms of a mixing angle $\theta$ ($c_\theta\equiv\cos\theta$, $s_\theta\equiv\sin\theta$), and 
\bea  \Pi_{Z_L Z_L}(T) &=&\left[\frac23 g^2 c_{\rm W}^2+\frac{g^2}{6c_{\rm W}^2}(1-2s_{\rm W}^2c_{\rm W}^2)+\frac{g^2}{c_{\rm W}^2}\left(1-2s_{\rm W}^2+\frac83 s_{\rm W}^4\right)\right] T^2, \\
\Pi_{\gamma_L \gamma_L}(T) &=& \frac{11}{3} e^2 T^2, \\
  \Pi_{Z \gamma_L}(T) &=& \frac{11}{6}eg \frac{c_{\rm W}^2-s_{\rm W}^2}{c_{\rm W}} T^2 \eea
 where $c_{\rm W}\equiv \cos\theta_W$ and  $s_{\rm W}\equiv \sin\theta_W$.
$Z_2(h, T)$  was presented previously  in the $g_Y=0$ limit in~\cite{Bodeker:1993kj}. 
Here we also included the effect of $g_Y$. In this formula we only included the dominant contribution of the zero Matsubara modes of bosons:
in this approximation there are no corrections induced by the top-quark Yukawa coupling. 
 
 \medskip 
 
 Up to higher-order terms, the correction to the bounce action is given by the new term of eq.\eq{dK},
 evaluated on the bounce computed ignoring this term.
 We find that the bounce action changes at the few $\%$ level, see fig.~\ref{fig:EffectiveCoupling}.

We do not compute the effect of terms with more than 2 derivatives, but we estimate that they can give effects comparable
to that of corrections to the 2-derivative term.
Indeed, loop corrections give higher-order Higgs derivative terms, which can be large 
when the Higgs has a sizeable coupling to some other particle
not much heavier than the Higgs itself.
At zero temperature, all masses come from the Higgs vev: in the limit of a large vev the Higgs is relatively lighter than $t,W,Z$,
because its mass is controlled by the Higgs self-coupling $\lambda$, which runs to relatively small values at large energy.
As a consequence, at $T=0$ and large vev one has $m_h \ll m_{t,W,Z}$, so that higher-order derivative terms are suppressed.
At finite temperature the Higgs receives an extra thermal mass given by the larger $y_t,g_1, g_2$ couplings:
as a consequence all thermal masses are comparable, and higher-order derivative terms could be significant.

 \begin{figure}[t]
\begin{center}
  \includegraphics[width=1.\linewidth]{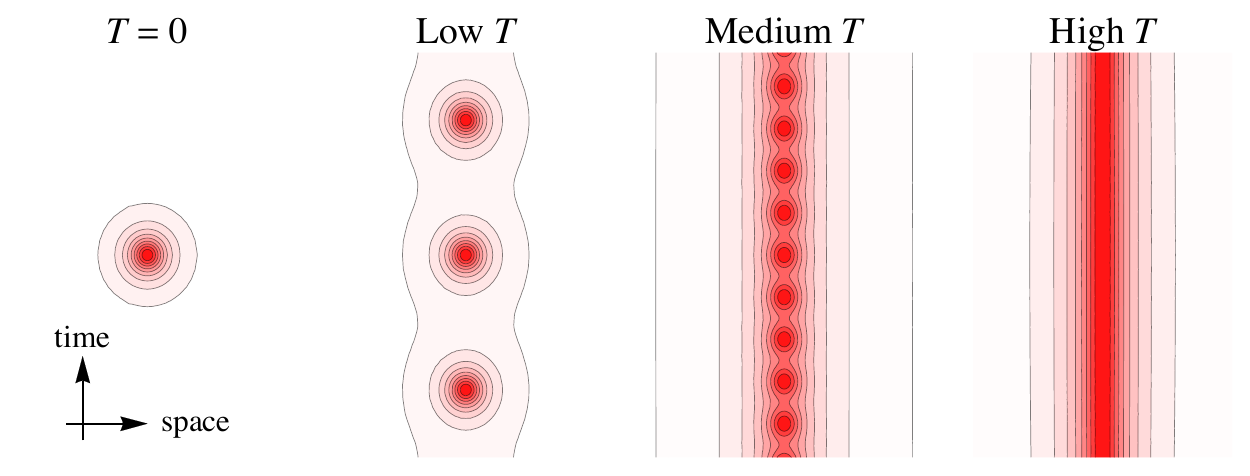}
  \end{center}
\caption{\em 
Bounces at different temperatures. The vertical axis represents the Euclidean time direction, and the horizonal axis represents the spatial radius.
At $T=0$ (left-most panel) the bounce solution enjoys an ${\rm O}(4)$ symmetry.
At finite temperature, the bounce solution becomes a series of bubbles
placed at distance $1/T$ in the time direction.
At large temperature (right-most panel) the bounce no longer depends on time.
 \label{fig:ThermalBounce} }
\end{figure}

\subsection{Is the thermal bounce time-independent?}\label{largeT}
The thermal tunneling rate at temperature $T$ is computed from the action 
\beq
S= 4\pi \int_0^{\beta}dt_E\int_{0}^{\infty}
dr~ r^2\left[
\frac{1}{2}\left(
\frac{\partial h}{\partial t_E}
\right)^2 + \frac{1}{2}\left(
\frac{\partial h}{\partial r}
\right)^2  + V_{\rm eff}(h)
\right] \eeq
of a bounce
$h(r,t_E)$ where $r \equiv \sqrt{|\vec{x}|^2}$ is the spatial radius and $t_E=-it$ the Euclidean time.
The bounce solves the classical equation
\begin{equation}\label{eq:BounceEquation}
\frac{\partial^2 h}{\partial t_E^2} + \frac{\partial^2 h}{\partial r^2} + \frac{2}{r}\frac{\partial h}{\partial r} = \frac{d V_{\rm eff}}{d h}
\end{equation}
with modified boundary conditions
\begin{equation}\label{eq:Boundaries}
\left.\frac{\partial h}{\partial t_E}\right|_{t_E = 0,\pm 1/2T} = 0,\qquad
\left.\frac{\partial h}{\partial r}\right|_{r = 0} = 0~,\qquad
\lim_{r \to \infty}h(r,t_E) = 0
\end{equation}
that impose periodicity in Euclidean time, $h(r, t_E + \beta) = h(r,t_E)$.

One trivial solution is a bounce constant in time, and normally this is the lowest-action solution at large enough temperature,
as illustrated in the right panel of fig.\fig{ThermalBounce} (see also~\cite{oldTM}).
Indeed, when a theory has a characteristic energy scale $m$, it sets the scale of the O(4)-symmetric bounce valid at $T=0$.
At low $T$ the periodicity is
irrelevant, because the time period is much longer than the scale of the $T=0$ bounce, 
as illustrated in the left panel of fig.\fig{ThermalBounce}.
For $T$ much larger than the scale of the $T=0$ bounce,
the short time periodicity implies that (if the vacuum instability still exits) the bounce 
becomes constant in time.
Thereby the action of the time-independent bounce 
scales as $S\propto 1/T$ and is  given by $S\sim m/T$, such that it dominates tunnelling above some critical temperature of order $m$.

Previous computations of thermal decay in the SM at $T\gg M_h$ assumed a time-independent thermal bounce.
However, the physical Higgs mass $M_h$ is not the relevant energy scale for the instability of the SM Higgs potential.
Rather, $M_h$ can be neglected, obtaining a quasi-scale-invariant action for the Higgs.
The assumption that $T$ is much larger than the energy scale of the problem
must be reconsidered, in view of the fact that the problem does not have a characteristic energy scale.

\medskip

In the thermal bath, $h$ acquires a thermal mass $m_T = \kappa T$.
Therefore, the large temperature limit $T\gg m_T$ would 
correspond to $\kappa \ll 1$ and would give a constant $S\sim m_T/T =\kappa$.
The SM predicts $\kappa\sim g\sim 0.4$, see eq.~(\ref{sys:thermal}) 
and fig.~\ref{fig:EffectiveCoupling}:
it is not much smaller than unity, potentially threatening the validity of  
usual computations that assume a time-independent thermal bounce.
In order to settle the issue,
we investigate whether time-dependent bounces have lower action.

\begin{figure}[t]
\begin{center}
  \includegraphics[width=.45\linewidth]{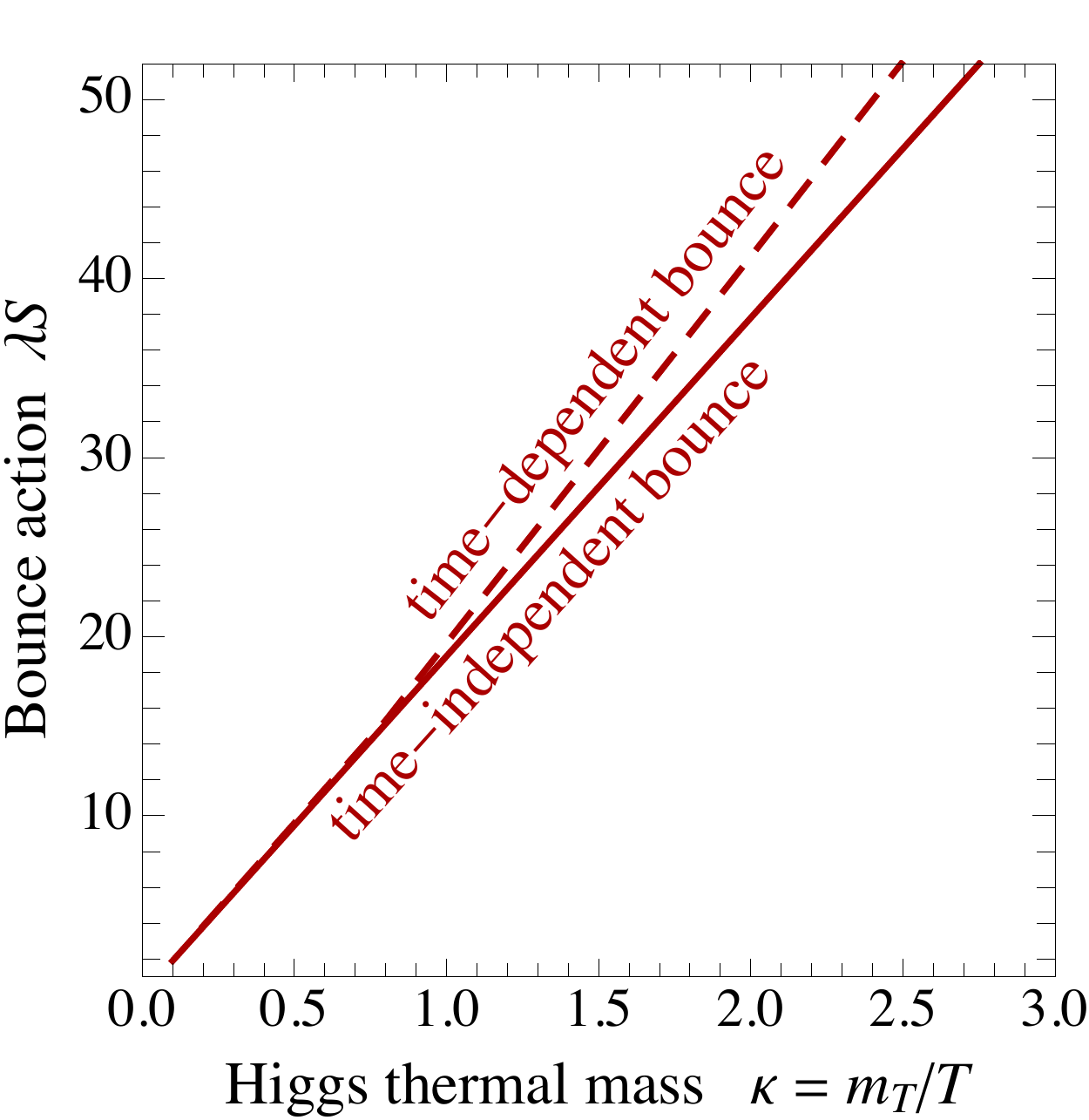}
  \end{center}
\caption{\em 
\label{fig:ComparisonT}
Rescaled thermal bounce action $\lambda S$ as a function  of the Higgs thermal mass $\kappa = m_T/T$.
The solid (dashed) line corresponds to the time-independent (time-dependent) bounce. }
\end{figure}

\medskip 

To start we consider a simplified SM-like potential $V_{\rm eff}(h) = \frac12 \kappa^2 T^2 h^2 -\frac14 \lambda h^4$ with
constant $\kappa$ and $\lambda$.
By rescaling $h(x)$ to a dimensionless $\eta(\xi)$ defined by
$h(x)=\eta(\xi) \kappa T/\sqrt{\lambda} $ and
$x_\mu = \xi_\mu/ \kappa T$ 
(we denote as $\tau$ and $\rho$ the dimensionless time and radius)
the action becomes
\beq
S= \frac{4\pi}{\lambda} \int_0^{\kappa}d\tau\int_0^{\infty}
d\rho ~\rho^2
\left[
\frac{1}{2}\left(
\frac{\partial \eta}{\partial\tau}
\right)^2 + \frac{1}{2}\left(
\frac{\partial \eta}{\partial \rho}
\right)^2  + \frac12 \eta^2 - \frac14 \eta^4
\right].
\eeq
This shows that $\lambda S$ does not depend on $\lambda$ and that
for the time-independent bounce, $\lambda S$ is proportional to $\kappa$. The precise result is
$\lambda S=6.015\pi \kappa$~\cite{Arnold}.
The rescaled action $\lambda S$ of a time-dependent bounce can be a more generic function of $\kappa$.
Figure~\ref{fig:ComparisonT} shows our numerical result for $\lambda S$, demonstrating that the time-dependent bounce
always has a higher action and is thereby subdominant.\footnote{Solving numerically the differential equation eq.~(\ref{eq:BounceEquation}) is not an easy task, since it is a non-linear equation with
non-trivial boundary conditions in space and time.
We discretise it on a space-time lattice, obtaining an ordinary non-linear equation $E_i = 0$ at each point $i$.
Next, we numerically minimise $\sum E_i^2$ applying the usual  Newton-like methods.
These need a starting ansatz, and
convergence is obtained provided that the starting point is good enough.
Appropriate choices are the O(4)-symmetric bounce,
or even the $T=0$ bounce of eq.\eq{h0}, provided that $h(0,0)$ is left as a free parameter.
Linear equations, such as boundary conditions, can be first imposed exactly, improving the procedure.
}

We next consider the full SM thermal potential:
the bounce action can significantly deviate from the above approximation,
but again the time-independent bounce dominates.

 \section{Conclusions}\label{concl}
 We reconsidered quantum and thermal vacuum decay in the SM.

 \medskip
 
Concerning vacuum decay, we validated the semi-analytical low-energy approximation for gravitational corrections
at leading-order  in $1/\Mp$
proposed in~\cite{0712.0242} (and wrongly criticized in~\cite{1601.06963,1606.00849})
through numerical computations in a toy model (section~\ref{toy}) and in the SM
 (section~\ref{SM}).
We generalised~\cite{0712.0242} allowing for a non-minimal scalar coupling $-\frac12 f(h){\cal R}$ to the curvature ${\cal R}$
and found a simplified expression for the leading-order gravitational correction to the bounce action
\beq \label{eq:Gexp3}
\Delta S_{\rm gravity}\simeq
 \frac{\pi^2}{6\bMp^2} \int dr~r^5  \bigg[\frac{h_0^{\prime 2}}{2} - V(h_0) - \frac{3}{r} f'(h_0)  h_0'
\bigg]^2 \ge 0
\eeq 
which makes clear that gravity suppresses Minkowski vacuum decay.
Going beyond this leading-order approximation we discussed how theories of quantum gravity can affect the result:
string models can give a landscape of new vacua,
agravity reduces the gravitational correction.

\medskip

The expansion parameter of thermal corrections is $g/\pi \sim 10^{-1}$ (larger than 
the expansion parameter $g^2/(4\pi)^2 \sim 10^{-3}$ of quantum corrections at $T=0$).
We found that 2-loop corrections to the thermal potential
and one-loop thermal corrections to the Higgs kinetic term 
change the bounce action by a small amount, at the few  $\%$ level,
as illustrated in fig.~\ref{fig:EffectiveCoupling}.
The SM meta-stability boundary in the ($M_t,M_h$) plane
 gets shifted by $+0.1\GeV$ in $M_t$ by a $3\%$ increase in the thermal bounce action $S$.
Taking into account that the two new effects that we added have opposite sign, 
fig.\fig{PhaseDiagram} shows the minor shift in the boundary, computed assuming
a reheating temperature of $10^{16}\GeV$.
Furthermore, we verified that the usual time-independent thermal bounce 
dominates over time-dependent bounces: 
this generically happens at large temperatures
but was not guaranteed within the SM, given that it is quasi-scale-invariant.

In conclusion, the residual theoretical uncertainty on SM meta-stability bounds is safely smaller than the 
experimental uncertainty, dominated by the uncertainty on the top mass.

\footnotesize

\subsection*{Acknowledgements}
We thank J.R. Espinosa, M. Garny,  M. Quiros and V. Rychkov for useful discussions. This work was supported by the grant 669668 -- NEO-NAT -- ERC-AdG-2014.

\appendix

\end{document}